\documentclass[11pt]{article}
\usepackage{amssymb}
\usepackage{epsf}
\begin{document}
\renewcommand{\thefootnote}{\fnsymbol{footnote}}
\renewcommand{\theequation}{{\rm\thesection.\arabic{equation}}}
\newcommand{\ee}{{{\eta}\over {2}}}
\newcommand{\beq}{\begin{equation}}
\newcommand{\eeq}{\end{equation}}
\newcommand{\beqa}{\begin{eqnarray}}
\newcommand{\eeqa}{\end{eqnarray}}
\newcommand{\beqan}{\begin{eqnarray*}}
\newcommand{\eeqan}{\end{eqnarray*}}
\newcommand{\half}{{{1}\over{2}}}
\newcommand{\ihalf}{{{i}\over{2}}}
\newcommand{\quar}{{{1}\over{4}}}
\newcommand{\la}{\lambda}
\newcommand{\La}{\Lambda}
\newcommand{\si}{\sigma}
\newcommand{\Si}{\Sigma}
\newcommand{\tf}{\Theta}
\newcommand{\bra}{\langle 0|}
\newcommand{\ket}{|0\rangle}
\newcommand{\id}{{1}\hspace{-0.3em}{\rm{I}}}
\def\b{\beta}
\def\al{\alpha}
\def\m{\mu}
\def\l{\lambda}
\def\d{\delta}
\def\g{\gamma}
\def\a{\alpha}
\def\t{\tau}
\def\e{\epsilon}
\def\r{\rho}
\def\d{\delta}
\def\s{\sigma}
\def\be{\begin{equation}}
\def\ee{\end{equation}}
\renewcommand{\ldots}{...}
\begin{flushright}BN--TH--2001--01\\
\end{flushright}
\vskip.5em
\begin{center}
{\Large{\bf A New Basis For Bethe Vectors Of The Heisenberg Model}}\vskip1.5em
T.-D. Albert\footnote{Talk presented by T.--D. Albert at the NATO workshop ``Dynamical Symmetries of Integrable Quantum Field 
Theories and Lattice Models'', Kiev 2000} and K. Ruhlig\\
{\bf(\{t-albert, ruhlig\}@th.physik.uni-bonn.de)}\\
{\sl\small{Physikalisches Institut Universit\"at Bonn, Nu{\ss}allee 12, D-53115 Bonn, Germany}}
\end{center}
\begin{abstract}
\noindent We present a detailed construction of a completely symmetric representation of the monodromy matrix by the use of Drinfel'd twists for the rational $sl(3)$  Heisenberg model without refering to the special symmetry of the model. With the help of this representation we are able to resolve the hierarchy of the nested Bethe wavevectors for the $sl(3)$ invariant rational Heisenberg model.
\end{abstract}
\vspace{-.4cm}\section{Introduction}\setcounter{equation}{0}\vspace{-.1cm}
Despite the indisputable achievements of the Quantum Inverse Scattering Method (QISM) and the rather simple action of the inverse problem operators, which can be interpreted as creation and annihilation operators for quasiparticles and as generating functions for the conserved quantities respectively, the study of correlation functions and formfactors has proven to be rather intricate. This is partly due to the fact that the solution of the inverse problem (expressing the original microscopic operators by means of the operators figuring in the algebraic Bethe ansatz) has only been achieved recently \cite{maillet, qis1, qis2} for some particular models tractable by the algebraic Bethe ansatz. An important contribution has been the application of factorizing Drinfel'd twists by Maillet and Sanchez de Santos to inhomogeneous spin chain models \cite{ms} for which the algebraic Bethe ansatz is available. It was shown that the similarity transformation provided by the factorizing twists gives rise to a completely symmetric representation of the monodromy matrix. They used as paradigmata for their argumentation the rational XXX and the trigonometric XXZ-model with underlying group $sl(2)$. A striking aspect of the results in \cite{ms} is related to the fact that  no polarization clouds are attached to quasiparticle creation and annihilation operators in the basis in which the monodromy matrix is completely symmetric. This means in terms of a particle notation that no virtual particle--antiparticle pairs are present in the wave vectors generated by the action of the creation operators to the ground state (the reference state of the Bethe ansatz), or in spin chain terminology that the creation  and annihilation  operators  are exclusively  built from local spin 
raising and spin lowering operators respectively (that is,  there are no compensating pairs of local raising and lowering spin operators). It was noted in \cite{ms}  that this latter feature underscores the neat connection between the quantum spin chain 
models and their respective quasiclassical limits, which are  Gaudin magnets, insofar as the appearance of the quasiparticle operators of the quantum models 
in the particular basis  differ from the corresponding operators in the quasiclassical limit models  only by a ``diagonal dressing'' (see below). A generalized transformation has subsequently been used to resolve the nested hierarchy in the Bethe vectors of the $sl(n)$ XXX-model \cite{abfr}.\\
In the spirit of a more pedagogicial introduction this technique is demonstrated in the case of the $sl(3)$ rational Heisenberg model, allowing a more detailed derivation. The generalized results valid for $sl(n)$ can be obtained straightforwardly.\\
The plan of  these notes is as follows: Section 2 sets the notation, while section 3 explains nested Bethe ansatz. In section 4 we present the construction of the factorizing twist and the induced similarity transformation for the creation operators. Section 5 is devoted to the resolution of the Bethe hierarchy. Section 6 contains our conclusions. 
\vspace{-.4cm}\section{Basic definitions and notation}\setcounter{equation}{0}\vspace{-.1cm}
The object central to the QISM is the so called {\it R-matrix}, which encodes the dynamical symmetry of the model at hand.\\
Taking $\{e_i\}$ as the standard orthonormal basis of $V\cong\mathbb{C}^n$, this R-matrix gives a representation of the Boltzmann weights $R_{ij}^{kl}(\la)$ for a vertex of a two-dimensional lattice with spin variables $j\in sl(n)$ placed on the bonds via the map $R:V_1\otimes V_2\;\;\rightarrow\;\;V_1\otimes V_2$ on the basis $e_i\otimes e_j$\footnote{The tensor product of two $n\times n$ matrices $X,\,Y$ is defined as $\left(X\otimes Y\right)_{kl}^{ij}=X_{ik}\,Y_{jl}\quad(i,j,k,l=1,\ldots,n)$ and the matrix product of two $n^2\times n^2$ matrices $S,T$ in this notation is given by $\left(ST\right)_{kl}^{ij}=S_{mn}^{ij}T_{kl}^{mn}$.}
\beqan
R(e_i\otimes e_j)=\sum_{k,l}(e_k\otimes e_l)R_{ij}^{kl}\;.
\eeqan
The R-matrix depends on two spectral parameters $\la_1$ and $\la_2$ associated to these two vector spaces. In what follows we denote it by $R_{12}(\la_1,\la_2)$.\\ 
In the case at hand the R-matrix is given by
\beq
R_{ij}(\l_i,\l_j)=b(\l_i,\l_j)\id_{ij}+c(\l_i,\l_j)\Pi_{i\,j},
\eeq
i.e., the matrix elements $R_{ij}$ are 
\beqan
\left(R_{ij}(\l)\right)^{\a_i\a_{j}}_{\b_i\b_{j}}=b(\l)\d_{\a_i\b_{i}}\d_{\a_{j}\b_j}+c(\l)\d_{\a_i\b_{j}}\d_{\a_{j}\b_i}
,\label{R}
\eeqan
with 
\beq
b(\l,\m)={{\l-\m}\over{\l-\m+\eta}},\quad c(\l,\m)={{\eta}\over{\l-\m+\eta}},\quad\eta\in\mathbb{C}\label{b,c}\,.
\eeq
Identifying one of the two linear spaces of this R-matrix with the Hilbert space ${\cal{H}}_i$ of the spin corresponding to site $i$ of a linear chain, we construct the quantum {\sl Lax operator}
\beq
L_i(\la,z_i)=R_{0i}(\la,z_i)\label{laxop}
\eeq
where $z_i$ is an arbitrary inhomogeneity parameter depending on site $i$. The Lax operator acts in the tensorproduct $V_0\otimes{\cal H}_i$, where the auxilliary space $V_0$ is isomorphic to $\mathbb{C}^n$.\\
The quantum {\sl monodromy matrix} for a chain of length $N$ is obtained as an ordered product of Lax operators\footnote{Sometimes we just omit the dependence on the local inhomogeneities $\{z_i\}$ and on the dimension $n$ and $N$, writing $T(\la)$.} 
\beq
T_N^{(n)}(\la,\{z_1,\ldots,z_N\})=L_{N}(\la,z_N)\,L_{N-1}(\la,z_{N-1})\ldots L_{1}(\la,z_1)\;.\label{monodromy}
\eeq
It can be represented in the auxilliary space as a $n\times n$ matrix whose elements are linear (nonlocal, i.e. acting on all sites $i=1,\ldots,N$) operators acting in the quantum space of states of the chain ${\cal{H}}=\otimes_{i=1}^N{\cal{H}}_i$\,:
\beq
T_N^{(n)}(\la)=\left(\matrix{
A_{11}&\ldots&A_{1\,n-1}&B_{1}\cr
\vdots&\ddots&\vdots&\vdots\cr
A_{n-1\,1}&\ldots&A_{n-1\,n-1}&B_{n-1}\cr
C_1&\ldots&C_{n-1}&D\cr}\right).\label{monodromymatrix}
\eeq
By taking the trace in the auxilliary space we obtain the {\sl transfer matrix}
\beq
{\cal{F}}^{(n)}(\la)=tr_0T^{(n)}(\la)=\sum_{i=1}^{n-1}A_{ii}(\la)+D(\la)\label{transfer}
\eeq
which can be interpreted as a discrete evolution operator for one time step, being related to the {\sl Hamiltonian} of the quantum chain by 
\beq
H=i{{d\,\ln{\cal F}(\la)}\over{d\,\la}}\Big|_{\la=0}\;.\label{ham}
\eeq
The partition function of the original vertex model being given as the sum over all possible spin configurations $\cal{C}$ weighted by the energy of the respective configuration can thus be reformulated in terms of the transfer matrix
\beqan
{\cal{Z}}=\sum_{{\cal{C}}}\exp\{E({\cal{C}})\}=Tr_{\cal{H}}({\cal{F}}^N)
\eeqan
where the trace $Tr_{\cal{H}}$ is taken in the whole Hilbert space of states.\\
This reduces the problem of calculating the partition function to the diagonalization of an operator in $\cal{H}$.\par\noindent
The {\sl integrability} of the model is based on the fact that a local commutation relation for the Lax operators exists ($\tilde{R}=\Pi R$)
\beq
\tilde{R}_{12}(\la,\mu)\,L_{i}(\la,z_i)\otimes L_{i}(\mu,z_i)
= L_{i}(\mu,z_i)\otimes L_{i}(\la,z_i)\,\tilde{R}_{12}(\la,\mu)\label{lcr}
\eeq 
where the R-matrix acts in the tensorproduct of two auxilliary spaces denoted $1$ and $2$.
The local relation (\ref{lcr}) induces a global relation for the monodromy matrix\index{monodromy matrix}\footnote{Note that the difference property of the R-matrix stipulates the distribution of the local inhomogeneities $\{z_i\}$ to be identical.}
\beq
\tilde{R}_{12}(\la,\mu)\,T(\la,\{z_i\})\otimes T(\mu,\{z_i\})
=T(\mu,\{z_i\})\otimes T(\la,\{z_i\})\,\tilde{R}_{12}(\la,\mu)\;.\label{gcr}
\eeq 
Taking the trace in the auxilliary spaces and using the cyclic property of the trace we obtain for the transfer matrix (\ref{transfer})  ($trA\otimes B=trA\,trB$)
\beq
\left[{\cal F}(\la),{\cal F}(\mu)\right]=0
\eeq
i.e. we have a one parameter family of commuting transfer matrices. It follows that the model is integrable in the generalized Liouville sense.\\
The conserved quantities 
\beq
I_k=i^k{{d^k\,\ln{\cal F}(\la)}\over{d\,\la^k}}\Big|_{\la=0}\label{cq}
\eeq
couple ($k+1$) nearest neighbors on the chain, with $I_1$ being the Hamiltonian (cf. (\ref{ham})) with nearest neighbor interaction.\\
The consistency of  (\ref{gcr}) (i.e. the associativity of a product of three operators $T$) implies the relation to be fulfilled by the R-matrix
\beqa
&&R_{12}(\la_1,\la_2)R_{13}(\la_1,\la_3)R_{23}(\la_2,\la_3)\nonumber\\
&&\hspace{5em}=R_{23}(\la_2,\la_3)R_{13}(\la_1,\la_3)R_{12}(\la_1,\la_2)
\label{ybe}
\eeqa
on the tensor product of three auxilliary spaces. $R_{ij}$ is acting nontrivially only on the space $V_i$ and $V_j$. \\
This relation is commonly denoted as {\sl Yang-Baxter equation}. It can be considered as the hallmark of integrability.\\
As can be seen from (\ref{gcr}), for $\la=\mu$, the R-matrix\index{R-matrix} reduces to the permutation operator $\Pi$ acting on the auxilliary spaces $V_1\otimes V_2$
\beq
R_{12}(0)=\Pi_{12}\label{regular}\;.
\eeq
It is furthermore unitary: $R_{ij}(\la_i,\la_j)R_{ji}(\la_j,\la_i)=\id$.\\
With the notation $T_{0,23}=R_{03}R_{02},\;\;R_{0i}\equiv R_{0i}(z_i)$,  where the index $0$ refers to an  auxilliary 
space $\mathbb{C}^n_{(0)}$, one may rewrite Eq. (\ref{ybe}) in the form $R^{\s_{23}}_{23}T_{0,23}=T_{0,32}R^{\s_{23}}_{23}$
with $\s_{23}$ the transposition of space labels $(2,3)$.\\
With the definition $T_{0,1\ldots N}=R_{0N}\ldots R_{01}$ it is straightforward to generalize this to a N-fold tensor product of spaces
\beq
R^{\sigma}_{1\ldots N}T_{0,1\ldots N}=T_{0,\sigma(1)\ldots \sigma(N)}R^{\sigma}_{1\ldots N} \label{lcrperm}
\eeq
where $\sigma$ is now an element of the symmetric group ${\cal{S}}_N$ and $R^{\sigma}_{1\ldots N}$ 
denotes a product of elementary $R$-matrices, the product corresponding to a decomposition of $\sigma$ into elementary transpositions. \\
The order of the upper matrix indices $\a_i$ of the $R^{\s}$ reads according to the above prescription as follows $\left(R^{\s}_{1\ldots N}\right)^{\a_{\s(N)}\ldots\a_{\s(1)}}_{\quad\b_{N}\ldots\b_{1}}$.\\
Eq. (\ref{lcrperm}) implies the {\sl composition law}
\beq
R^{\sigma'\sigma}_{1\ldots N}=R^{\sigma}_{\sigma'(1)\ldots \sigma'(N)}  R^{\sigma'}_{1\ldots N}\label{complaw} 
\eeq
for a product of two elements in ${\cal{S}}_N$. 
\vspace{-.4cm}\section{The Nested Bethe Ansatz}\setcounter{equation}{0}\vspace{-.1cm}
The diagonalization of ${\cal{F}}^{(3)}(\la)$ \cite{kulresch} for all values of $\la$ starts with the existence of a pseudovacuum $\Omega$.\footnote{This pseudovacuum coincides with the physical vacuum only for the ferromagnetic case, which is why it is also denoted reference state.}
For this purpose one introduces a lowest weight state with respect to $sl(3)$ (local vacuum) $v^{(i)}_3\in {\cal H}_i$:
\beq
v^{(i)}_3={}^t\left(0,0,1\right).
\eeq
The action of the Lax operator is given by
\beqa
(L_i(\l,z_i))_{33}v^{(i)}_3&=&v^{(i)}_3,\nonumber\\
(L_i(\l,z_i))_{kk}v^{(i)}_3&=&b(\l,z_i)v^{(i)}_3\quad\mbox{for }k\neq 3, \nonumber\\
(L_i(\l,z_i))_{3k}v^{(i)}_3&=&c(\l,z_i)E^{(i)}_{k,3}v^{(i)}_3\quad\mbox{for }k< 3\,,\label{Lv}
\eeqa
with $\left(E_{ab}\right)^m_n=\delta_{a,m}\delta_{b,n}$ the root of $sl(3)$. All other operator entries annihilate the local vacuum.\\
A global vacuum for the operators of the monodromy matrix can be constructed as follows
\beq
\Omega^{(3)}_N=\otimes^N_{i=1}v^{(i)}_3.\label{vac}
\eeq
Using the expression of the monodromy matrix in the form (\ref{monodromymatrix}) specialized to $n=3$
we obtain 
\beqa
A_{ab}(\l)\Omega^{(3)}_N&=&\d_{ab}\prod_{i=1}^N b(\l,z_i)\Omega^{(3)}_N\label{AO}\,,\;\;\;D(\l)\Omega^{(3)}_N=\Omega^{(3)}_N,\nonumber\\
B_a(\l)\Omega^{(3)}_N&=&0.\label{evvac}
\eeqa
The operators $B_a(\l)$ can be interpreted as annihilation operators while $C_a(\l)$ act as creation operators.\\
In order to obtain eigenstates to the transfer matrix (\ref{transfer}) we will consider a vector which consists of linear superpositions of products of $C_i$ (with $card\,\left\{C_i\right\}$ fixed)
\beq
\Psi^{(3)}=\sum_{\al_1,\ldots,\al_p}\Phi^{(2)}_{\al_1,\ldots,\al_p}\,C_{\al_1}(\la_1)\ldots C_{\al_p}(\la_p)\,\Omega ^{(3)}_N\label{nbv}
\eeq
with c-number coefficients $\Phi_{\al_1,\ldots,\al_p}^{(2)}$ still to be determined.\\
The eigenvalues of the transfermatrix can be obtained by commuting the operators $A_{ii}(\l)$ and $D(\l)$ through to the vacuum, for which the eigenvalues of the transfer matrix are already known by (\ref{evvac}).\\
From the global commutation relation (\ref{gcr}) we obtain 
\beqa
A_{ab}(\l)C_c(\m)&=&{(R^{(2)}(\l,\m))^{dp}_{bc}\over b(\l,\m)}C_p(\m)A_{ad}(\l)-{c(\l,\m)\over b(\l,\m)}C_b(\l)A_{ac}(\m),\nonumber\\
D(\l)C_a(\m)&=&{1\over b(\m,\l)}C_a(\m)D(\l)-{c(\m,\l)\over b(\m,\l)}C_a(\l)D(\m)\;.\label{acd}
\eeqa
The appearance of the $sl(2)$ R-matrix $R^{(2)}_{ij}(\l)=b(\l)\id^{(2)}_{ij}+c(\l)\Pi^{(2)}_{i\,j}$ acting in $\mathbb{C}^2\otimes\mathbb{C}^2$ in the commutator relations (\ref{acd}) constitutes the main difference as compared to the $sl(2)$ XXX-model.
By inspection of (\ref{acd}) we notice that only the first term on the r.h.s. can produce eigenvectors, while the contribution of the second {which exchanges the arguments} must vanish, giving  rise to equations to be fulfilled by the parameters $\{\la_i\}$, the Bethe ansatz equations.\\
The action of  $D(\l)$ on (\ref{nbv}) yields for the wanted term
\beqan
\prod_{i=1}^m {1\over b(\l_i,\l)}\Phi^{(2)}_{\a_1\ldots\a_m}C_{\a_1}(\l_1)\ldots C_{\a_m}(\l_m)\Omega^{(3)}_N.\label{D3eigen}
\eeqan
All other terms are unwanted.
The action of $A_{aa}(\l)$ (summation over $a=1,2$) on (\ref{nbv}) yields in turn
\beqan
&&\prod_{i=1}^Nb(\l,z_i)\prod_{j=1}^m{1\over b(\l,\l_j)}\;\Phi^{(2)}_{\a_1\ldots\a_m}C_{p_1}(\l_1)\ldots C_{p_m}(\l_m)\Omega^{(3)}_N\nonumber\\
\times&&\hspace{-1em}(R^{(2)}(\l,\l_1))^{d_1\,p_1}_{a\,\a_1}(R^{(2)}(\l,\l_2))^{d_2\,p_2}_{d_1\,\a_2}\ldots (R^{(2)}(\l,\l_{m}))^{a\quad\;\; p_{m}}_{d_{m-1}\,\a_{m}}.\label{A3eigen}\nonumber\\
\eeqan
This term results in an eigenvector for $A_{aa}(\l)$ if the following additional relation is fulfilled:
\beq
(R^{(2)}(\l,\l_1))^{d_1\,p_1}_{a\,\a_1}\ldots(R^{(2)}(\l,\l_{m}))^{a\,p_{m}}_{d_{m-1}\,\a_{m}}\Phi^{(2)}_{\a_1\ldots\a_m}=f^{(2)}(\l) \Phi^{(2)}_{p_1\ldots p_m}.\label{Rphi}
\eeq
This can be reformulated with the help of the definitions 
\beqan
L_i^{(2)}(\l)&=&R^{(2)}_{0i}(\l),\;\;T^{(2)}_m(\l)=L_m^{(2)}(\l,\l_m)\ldots L_1^{(2)}(\l,\l_1),\nonumber\\
{\cal{F}}^{(2)}(\l)&=&tr_0T^{(2)}_m(\l)\label{def2}
\eeqan
to give
\beq
{\cal{F}}^{(2)}(\l)\Phi^{(2)}=f^{(2)}(\l) \Phi^{(2)}\quad\mbox{with }\Phi^{(2)}\in{\cal H}^{(2)}=(\mathbb{C}^2)^{\otimes m}.\label{eigen2}
\eeq
We thus have obtained a similar problem as before, but now for a $sl(2)$ chain of length $m$, with the inhomogenities given by the parameters $\l_1,\ldots,\l_m$ of the $sl(3)$ problem.
\\
This fact inspires us to choose the coefficient $\Phi^{(2)}$ analogously to the $sl(3)$ case
\beq
\Phi^{(2)}=C^{(2)}(\l^{(2)}_1)\ldots C^{(2)}(\l^{(2)}_q)\Omega^{(2)}_m.\label{phi2}
\eeq
The vacuum is given by
\beq
\Omega^{(2)}_m=\otimes_{i=1}^mv^{(i)}_2,\quad\mbox{with }v^{(i)}_2=\left(\begin{array}{c}0\\1\end{array}\right)\in {\cal H}_i^{(2)}=\mathbb{C}^2
\eeq
and the operators of the $2\times 2$ monodromy matrix fulfill
\beqa
A^{(2)}(\l)\Omega^{(2)}_m&=&\prod_{i=1}^mb(\l,\l_i)\Omega^{(2)}_m,\;\;\;D^{(2)}(\l)\Omega^{(2)}_m=\Omega^{(2)}_m,\nonumber\\
B^{(2)}(\l)\Omega^{(2)}_m&=&0.
\eeqa
The commutation relations are now 
\beqa
A^{(2)}(\l)C^{(2)}(\m)&=&{1\over b(\l,\m)}C^{(2)}(\m)A^{(2)}(\l)-{c(\l,\m)\over b(\l,\m)}C^{(2)}(\l)A^{(2)}(\m)\label{Akom2},\nonumber\\
D^{(2)}(\l)C^{(2)}(\m)&=&{1\over b(\m,\l)}C^{(2)}(\m)D^{(2)}(\l)-{c(\m,\l)\over b(\m,\l)}C^{(2)}(\l)D^{(2)}(\m),\nonumber\\\label{Dkom2}
[C^{(2)}(\l),C^{(2)}(\m)]&=&0\,.\label{CC2}
\eeqa
The eigenvalue $f^{(2)}$ in (\ref{eigen2}) is then given by
\beqan
f^{(2)}(\l)=\prod_{i=1}^m b(\l,\l_i)\prod_{j=1}^q{1\over b(\l,\l^{(2)}_j)}+\prod_{j=1}^q{1\over b(\l^{(2)}_j,\l)}\,.\label{eigenw2}
\eeqan
The eigenvalue of the original $sl(3)$ case finally is 
\beqan
&&{\cal{F}}^{(3)}(\l)\Psi_3(\l_1,\ldots,\l_m;\Phi^{(2)})=f^{(3)}(\l)\Psi_3(\l_1,\ldots,\l_m;\Phi^{(2)})\nonumber\\
&&f^{(3)}(\l)=\prod_{i=1}^m {1\over b(\l_i,\l)}+\nonumber\\
&&+\prod_{i=1}^N b(\l,z_i)\prod_{j=1}^m{1\over b(\l,\l_j)}\left\{\prod_{i=1}^m b(\l,\l_i)\prod_{j=1}^q{1\over b(\l,\l^{(2)}_j)}+\prod_{j=1}^q{1\over b(\l^{(2)}_j,\l)}\right\}.
\eeqan
Summarizing, $\Psi_3$ in the form (\ref{nbv}) is eigenvector of the transfer matrix ${\cal{F}}^{(3)}(\l)=\sum_{i}T_{ii}(\l)$ if\\
\hspace*{.8cm} $i)$ the parameters $\l_1,\ldots,\l_p$ satisfy a certain system of equations, the famous Bethe ansatz equations\\
and if \\
\hspace*{.8cm} $ ii)$ the c-number coefficients are chosen s.t. they constitute the components of a inhomogeneous $sl(2)$ transfer matrix, with the inhomogeneities given by the spectral parameters of the $sl(3)$ Bethe vector.\\\\
The Bethe ansatz equation become very intricate as the spectral parameters of the reduced ($sl(2)$) problem have to fulfill similar equations, thus yielding a set of coupled algebraic equations determining the set of numbers $\la_{j}$. They can be derived by the following trick: The general eigenvalue $f(\la)$ should be analytical in $\la$ as it is to generate the eigenvalues of the conserved quantities through (\ref{cq}). Consequently the residue at singular points is to vanish, which yields exactly the Bethe equations which in turn assure the vanishing of the unwanted terms.\\
We now want to show that the hierarchical wavefunction (\ref{nbv})
\beqan
\Psi_3 (N;\l_1,\ldots,\l_{p})=\sum_{\a_1,\ldots,\a_{p}}
\Phi^{(2)}_{\a_1\ldots\a_{p}}
C_{\a_1}(\l_1)\ldots C_{\a_{p}}(\l_{p})\,\Omega^{(3)}_3
\eeqan
is invariant under permutations of the parameters $\l_i$ \cite{takh}. In the  $sl(2)$ case this is obviuos as the operators $C(\l)$ commute for all spectral parameters (\ref{CC2}), while for $sl(3)$ it follows from (\ref{gcr}) that $C_i$ and $C_k$ do not commute anymore:
\beq
C_i(\la)C_j(\mu)=C_k(\mu)C_l(\la)\left(\tilde{R}^{(2)}(\la,\mu)\right)_{ij}^{kl}\label{crc}\;.
\eeq
Exchanging the parameters $\l_i$ and $\l_{i+1}$ in (\ref{nbv}), we obtain (sum convention)
\beqa
\Psi_3^i (\l_1,\ldots,\l_{p})=
\Phi^{(2),i}_{\a_1\ldots\a_{p}}
C_{\a_1}(\l_1)\ldots C_{\a_i}(\l_{i+1})C_{\a_{i+1}}(\l_i)\ldots C_{\a_{p}}(\l_{p})\,\Omega^{(3)}_N.\nonumber\\\label{psii}
\eeqa
where the index $i$ in $\Phi^{(2),i}_{\a_1\ldots\a_{p}}$ indicates that  the monodromy matrix
\beqa
T^{(2),i}_{m}(\l)&=&L_{m}^{(2)}(\l,\l_{m})\ldots L_{i+1}^{(2)}(\l,\l_{i})L_{i}^{(2)}(\l,\l_{i+1})\ldots L_1^{(2)}(\l,\l_1)\nonumber\\
\label{moni}
\eeqa
occurs in the wavefunction. Commuting the operators $C_{\a_i}(\l_{i+1})$ and $C_{\a_{i+1}}(\l_i)$ in (\ref{psii}) we get with the help of (\ref{crc})
\beqan
&&\Psi_3^i (\l_1,\ldots,\l_{p})=\nonumber\\
&&\Phi^{(2),i}_{\a_1\ldots\a_{p}}
C_{\a_1}(\l_1)\ldots ( \tilde{R}^{(2)}(\l_{i+1},\l_i))^{mn}_{\a_i\a_{i+1}}C_{m}(\l_i)C_{n}(\l_{i+1})\ldots C_{\a_{p}}(\l_{p})\,\Omega^{(3)}_N.\label{psiiR}
\eeqan
The action of $\tilde{R}$ on $\Phi^{(2),i}_{\a_1\ldots\a_{m_1}}$ can be computed using the following relation which is a consequence of the YBE (\ref{ybe}):
\beqan
&&\tilde{R}^{(2)}_{i\,i+1}(\l_{i+1},\l_{i})L_{i+1}^{(2)}(\l,\l_{i})L_{i}^{(2)}(\l,\l_{i+1})=\nonumber\\
&&\hspace{4em}L_{i+1}^{(2)}(\l,\l_{i+1})L_{i}^{(2)}(\l,\l_{i}) \tilde{R}^{(2)}_{i\,i+1}(\l_{i+1},\l_{i})\,.
\eeqan
We now obtain a relation between (\ref{monodromy}) and (\ref{moni}):
\beq
T^{(2)}_{m}(\l) \tilde{R}^{(2)}_{i\,i+1}(\l_{i+1},\l_{i})= \tilde{R}^{(2)}_{i\,i+1}(\l_{i+1},\l_{i})T^{(2),i}_{m}(\l).
\eeq
As the coefficient $\Phi^{(2),i}_{\a_1\ldots\a_{m_1}}$ is given by (\ref{phi2}), we obtain with the help of $R_{ij}v_3^{(i)}\otimes v_3^{(j)}=v_3^{(i)}\otimes v_3^{(j)}$ the relation between $\Phi^{(2),i}_{\a_1\ldots\a_{m}}$ and $\Phi^{(2)}_{\a_1\ldots\a_{m}}$
\beq
\Phi^{(2)}_{\a_1\ldots\a_{m}}=( \tilde{R}^{(2)}(\l_{i+1},\l_i))_{\b_i\b_{i+1}}^{\a_i\,\a_{i+1}}\Phi^{(2),i}_{\a_1\ldots\b_i\,\b_{i+1}\ldots\a_{m}}.
\eeq
This finally yields the desired result, the wavefunction (\ref{nbv}) is invariant under permutations of the parameters $\l$. 
\vspace{-.4cm}\section{The F-Basis}\setcounter{equation}{0}\vspace{-.1cm}
Despite the achievements of the QISM\index{Quantum inverse scattering method (QISM)}, the treatmant of correlation functions proves to be rather intricate. The reason for this lies in the fact that the apparantly simple action of the quasiparticle  creation and annihilation operators which figure in the construction of the eigenfunctions (Bethe wavevectors\index{Bethe wavevector}) is obscured on the level of the microscopic variables (spin raising and lowering operators) by nonlocal effects arising from polarization clouds (compensating pairs of local raising and lowering spin operators).\\
This property can already be seen in the example of the $sl(2)$ XXX-model, where the creation operators have the following form when expressed in the original microscopic spin variables
\beq
C(\la)=\sum_{i=1}^N\s_i^+\Omega_i+\sum_{i\neq j\neq k}\s_i^+(\s_j^-\s_k^+)\Omega_{ijk}+{\mbox {terms of higher order}}
\eeq
where higher order means more compensating exchange terms $(\s_j^-\s_k^+)$, and the $\;\Omega_i,\;\Omega_{ijk}$ etc. are diagonal and act on all site but $i$; $i,j,k$ and so on.\\
In a seminal  paper  Maillet  and  Sanchez de Santos  \cite{ms} revealed an application of so called factorizing Drinfel'd twists\index{factorizing twist} $F$ being defined by $R_{ij}=F_{ij}F^{-1}_{ji}$ to obtain a basis for the $sl(2)$ XXX and XXZ-model which allows to express the  creation and annihilation operators in a completely symmetric way with the further advantage of being polarization free, that is being built from the respective quasiclassical Gaudin operators dressed diagonally, thus minimizing quantum effects for these operators.\\
In the new basis the operator $\tilde{C}(\la)\equiv F(\la)\,C(\la)F^{-1}(\la)$ consists of only $N$ terms (as compared to $2^N$ in the original basis) and is of the form
\beq
\tilde{C}(\la)=\sum_{i=1}^N\s_i^+\otimes_{j\ne i}
{\left(\matrix{
b(\l,z_j)b_{ij}^{-1}&0\cr
0\quad\quad&1\cr
}\right)}_{[j]}\;.
\eeq 
These results allowed a simple derivation of formulas (previously known, but involving so called auxilliary fields \cite{korepinnorm}) for the scalar products of Bethe vectors\index{Bethe wavevector} and for correlation functions \cite{maillet, spm}.\\
The factorizing $F$-matrix for $N$ sites ($N$ quantum spaces) is defined by the relation $R^{\sigma}_{1\ldots N}(\la)=F_{\sigma(1\ldots N)}^{-1}(\la)F_{1\ldots N}(\la)$.
The starting point of paper \cite{ms} is the Drinfel'd factorizing twists of the elementary $sl(2)$ $R$-matrix:
$R_{12}=F^{-1}_{21}F_{12}$
where $F_{12}$ is given by 
\beqan
F_{12}\; =\; \left(\matrix{
1&{\;0\;}&{\;0\;}&{\;0\;}\cr
0&{\;1\;}&{\;0\;}&{\;0\;}\cr
0&c(z_1,z_2)&b(z_1,z_2)&0\cr
0&{\;0\;}&{\;0\;}&{\;1\;}\cr}\right)_{[12]}.
\label{F12}
\eeqan
The generalization of this formula to the $sl(n)$ case is
of the form
\beq
F_{12}=\sum_{n\ge\a_2\ge\a_1}P_{\a_1}^{1}P_{\a_2}^{2}\,\id_{12}+\sum_{n\ge\a_1>\a_2}P_{\a_1}^{1}P_{\a_2}^{2}
R_{12}^{\s_{12}}\,.
\label{F12a}
\eeq
Here ${[P^{i}_{\a}]}_{k,l}=\d_{k,\a}\d_{l,\a}$ is the projector on the $\a$ component acting in $i$-th space.\\
Generalizing this factorization matrix to the $N$-site problem one has to satisfy 
at least three properties for the $F$-matrix\footnote{The F-matrix defined above is only determined up to multiplication with a completely symmetric, non-degenerate matrix.}\\
$\bullet$ {\bf{A}} $\quad$ factorization, that is  
\beq
F_{\s(1)\ldots\s(N)}(z_{\s(1)},\ldots,z_{\s(N)}) 
R^{\s}_{1\ldots N}(z_{1},\ldots,z_{N}) = F_{1\ldots N}(z_{1},\ldots,z_{N})
\label{factor}
\eeq
for any permutation  $\s\in {\cal{S}}_N$;\\
$\bullet$ {\bf{B}} $\quad$ lower-triangularity;\\
$\bullet$ {\bf{C}} $\quad$ non-degeneracy.\\\\
The following expression satisfies the properties {\bf A},{\bf B} and {\bf C}:
\beq
F_{1\ldots N} = \sum_{\s\in {\cal{S}}_N}\sum^{\quad\quad *}_{\a_{\s(1)}\ldots\a_{\s(N)}}
\prod_{i=1}^N P^{\s(i)}_{\a_{\s(i)}} R_{1\ldots N}^{\s}(z_{1},\ldots,z_{N})\, .
\label{F}
\eeq
The sum $\sum^{*}$ extends over indices $\a_i$, which fulfill the conditions
\beqa
&\a_{\s(i+1)}\ge\a_{\s(i)}\quad \mbox{for}\quad\s(i+1)>\s(i)&\nonumber\\
&\a_{\s(i+1)}>\a_{\s(i)}\quad \mbox{for}\quad\s(i+1)<\s(i)&,
\label{cond}
\eeqa
i.e., the equality sign holds if the original labels of the sites under consideration still have the same order after the permutation $\s$. This form of the F-matrix ensures down-triangularity and as the diagonal elements are non-vanishing, also non-degeneracy{\footnote{There exists an explicit form of the inverse of $F$, but as it is not needed in what follows we refrain from citing it.}}.\\
To proof factorization, we decompose the permutation $\s$ into elementary transpositions $\s_i$ $\s=\s_{1}...\s_{k}.\;$ $\s_i$ is the transposition of sites $i$ und $i+1$. The R-matrix $R^{\s}$ can be decomposed using (\ref{complaw}), which entails for the F-matrix
\beqan
&&F_{\s(1)...\s(N)}\; R_{1...N}^{\s}\nonumber\\
&=&F_{\s_1\s_2...\s_k(1,\ldots,N)}\;
R_{\s_1\s_2...\s_{k-1}(1,\ldots,N)}^{\s_k}\;
R_{\s_1\s_2...\s_{k-2}(1,\ldots,N)}^{\s_{k-1}}
\ldots
R_{1\ldots N}^{\s_1}\nonumber\\
&=&F_{\s_1\s_2...\s_{k-1}(1,\ldots,N)}\;
R_{\s_1\s_2...\s_{k-2}(1,\ldots,N)}^{\s_{k-1}}\;
R_{\s_1\s_2...\s_{k-3}(1,\ldots,N)}^{\s_{k-2}}
\ldots
R_{1\ldots N}^{\s_1}\nonumber\\
&=&\ldots\ldots\ldots=F_{\s_1(1,\ldots,N)}\;R_{1\ldots N}^{\s_1}\;=\;F_{1...N}\,,
\eeqan
thus enabling us to prove property {\bf A} by restricting ourselves to elementary transpositions $\s_i$, i.e. considering  $F_{1\ldots i+1\;i\ldots N}R^{\s_i}_{1\ldots N}$
\beqa
F_{1\ldots i+1\;i\ldots N}R^{\s_i}_{1\ldots N}
&=&F_{\s_i(1\ldots i\;i+1\ldots N)}R^{\s_i}_{1\ldots N}\nonumber\\
&=&\sum_{\s\in S_N}\sum^{\quad\quad *(i)}_{\a_{\s_i\s(1)}\ldots\a_{\s_i\s(N)}}
\prod_{j=1}^N P^{\s_i\s(j)}_{\a_{\s_i\s(j)}} R_{\s_i(1,\ldots,N)}^{\s}
R^{\s_i}_{1\ldots N}\nonumber\\
&=&\sum_{\s\in S_N}\sum^{\quad\quad *(i)}_{\a_{\s_i\s(1)}\ldots\a_{\s_i\s(N)}}
\prod_{j=1}^N P^{\s_i\s(j)}_{\a_{\s_i\s(j)}} 
R^{\s_i\s}_{1\ldots N}\, ,
\label{FR1}
\eeqa
with the sum $\sum^{*(i)}$ now restricted by the conditions 
\beqa
&\a_{\s_i\s(j+1)}\geq\a_{\s_i\s(j)}\quad \mbox{for}\quad\s(j+1)>\s(j)&\nonumber\\
&\a_{\s_i\s(j+1)}>\a_{\s_i\s(j)}\quad \mbox{for}\quad\s(j+1)<\s(j)&
\label{cond(1)}
\eeqa
as (\ref{cond}) refers to the sites already permuted ${\widetilde{j}}=\s_i(j)$.  Substituting $\tilde{\s}=\s_i\s$ in (\ref{FR1}) yields
\beqa
F_{1\ldots i+1\;i\ldots N}R^{\s_i}_{1\ldots N}&=&
\sum_{\tilde{\s}\in S_N}\sum^{\quad\quad *}_{\a_{\tilde{\s}(1)}\ldots\a_{\tilde{\s}(N)}}
\prod_{j=1}^N P^{\tilde{\s}(j)}_{\a_{\tilde{\s}(j)}} 
R^{\tilde{\s}}_{1\ldots N}
\label{FR2}
\eeqa
with the condition in $\sum^{*}$ 
\beqa
&\a_{\tilde{\s}(j+1)}\ge\a_{\tilde{\s}(j)}\quad \mbox{for}\quad\s_i\tilde{\s}(j+1)>\s_i\tilde{\s}(j)&\nonumber\\
&\a_{\tilde{\s}(j+1)}>\a_{\tilde{\s}(j)}\quad \mbox{for}\quad\s_i\tilde{\s}(j+1)<\s_i\tilde{\s}(j)&.
\label{cond2}
\eeqa
The conditions (\ref{cond}) and (\ref{cond2}) result in the same expression if $\s^{-1}(i)$ and $\s^{-1}(i+1)$ are not neighbored. If they are, i.e.
\beqa
\s^{-1}(i)=\s^{-1}(i+1)\pm 1\, ,\label{np}
\eeqa
the strict inequality sign renders the same result for (\ref{cond2}) and (\ref{cond}), while in the case of the equality sign, i.e. if $\a_{\widetilde{\s_{j+1}}}=\a_{\widetilde{\s_{j}}}$, the relation
\beq
P^i_{\a}P^j_{\a}R_{ij}=P^i_{\a}P^j_{\a}\id_{ij}\label{ppr}
\eeq
ensures that the additional transposition $\s_i$ in (\ref{FR2}) with regard to (\ref{F}) has no effect, which proves property {\bf A}{\footnote{A different proof emphasizing the geometric nature can be found in \cite{abfpr}.}}.
We are now in the positon to compute the matrix elements of the monodromy matrix in the basis provided by the F-matrix ${\tilde{T}}_{ij}=F_{1\ldots N} T_{ij} F_{1\ldots N}^{-1}$. We start illustrating this procedure in the case ${\tilde{D}=\tilde{T}}_{33}$, and then proceed to calculate the creation operators in the new basis. We will not rely on the symmetry of the $sl(3)$ XXX-model, which makes it possible to extend our results straightforwardly to the generalized XXZ-model by replacing the rational parametrization (\ref{b,c}) by its trigonometric counterpart.\\
To obtain ${\tilde{D}}=F_{1\ldots N}D F_{1\ldots N}^{-1}$ consider the expression ($\widetilde{i}=\s(i)$)
\beqa
F_{1\ldots N}T_{33}&=&\sum_{\s\in S_N}\sum^{\quad\quad *}_{\a_{\tilde{1}}\ldots\a_{\tilde{N}}}
\prod_{i=1}^N P^{\tilde{i}}_{\a_{\tilde{i}}} R_{1\ldots N}^{\s}P_3^0 T_{0,1\ldots N} P_3^0\nonumber\\
&=&\sum_{\s\in S_N}\sum^{\quad\quad *}_{\a_{\tilde{1}}\ldots\a_{\tilde{N}}}
\prod_{i=1}^N P^{\tilde{i}}_{\a_{\tilde{i}}}P_3^0 T_{0,\tilde{1}\ldots \tilde{N}} P_3^0
 R_{1\ldots N}^{\s}\,.\label{Tnn1}
\eeqa
The projectors $P_3^0$ yield the element $T_{33}$. The second step involved (\ref{lcrperm}) and the fact that $R^{\s}_{1\ldots N}$ commutes with $P_3^0$. We now split the sum $\sum^*$ according to the number $k$ of the occurence of the index $3$:
\beqa
&&F_{1\ldots N}T_{33}\nonumber\\
&&=\sum_{\s\in S_N}\sum_{k=0}^N\sum^{\quad\quad *'}_{\a_{\tilde{1}}\ldots\a_{\tilde{N}}}
\prod_{j=N-k+1}^N\delta_{\a_{\tilde{j}},3}P_3^{\tilde{j}}\prod_{j=1}^{N-k} P^{\tilde{j}}_{\a_{\tilde{j}}}P_3^0 T_{0,\tilde{1}\ldots \tilde{N}} P_3^0
 R_{1\ldots N}^{\s}\,.\nonumber\\\label{Tnn2}
\eeqa
Considering a term of arbitrary but fixed multiplicity of $3$ in (\ref{Tnn2}), we get 
\beqa
&&\prod_{j=1}^{N-k} P^{\tilde{j}}_{\a_{\tilde{j}}}\prod_{j=N-k+1}^N P_3^{\tilde{j}} P_3^0\,T_{0,\tilde{1}\ldots \tilde{N}}\,P_3^0\nonumber\\
&=&\prod_{j=1}^{N-k} P^{\tilde{j}}_{\a_{\tilde{j}}}\prod_{N-k+1}^{N}\left(R_{0,\tilde{j}}\right)_{33}^{33}\,P_3^0\,T_{0,\tilde{1}\ldots \widetilde{N-k}}\,P_3^0\prod_{j=N-k+1}^N P_3^{\tilde{j}}\nonumber\\
&=&\prod_{j=1}^{N-k} P^{\tilde{j}}_{\a_{\tilde{j}}}P_3^0\,T_{0,\tilde{1}\ldots \widetilde{N-k}}\,P_3^0 \prod_{j=N-k+1}^N P_3^{\tilde{j}}\nonumber\\
&=&\prod_{i=1}^{N-k}\left(R_{0\tilde{i}}\right)_{3,\a_{\tilde{i}}}^{3,\a_{\tilde{i}}}\prod_{j=1}^{N-k} P^{\tilde{j}}_{\a_{\tilde{j}}}\prod_{j=N-k+1}^N P_3^{\tilde{j}}P_3^0\nonumber\\
&=&\otimes_{i=1}^N \mbox{diag}\{b(\l,z_i),b(\l,z_i),1\}\hspace{-.1em}\prod_{j=1}^{N-k} P^{\tilde{j}}_{\a_{\tilde{j}}}\hspace{-.5em}\prod_{j=N-k+1}^N \hspace{-.5em}P_3^{\tilde{j}}P_3^0\, .\label{prp}
\eeqa
We again used the identity $P^i_{\a}P^j_{\a}R_{ij}=P^i_{\a}P^j_{\a}\id_{ij}$ and the fact that the upper and lower indices of the R-matrix are connected by a permutation, and finally that $\a_{\tilde{i}}\neq 3$ for $i\leq N-k$. In the last line we inserted $\left(R_{0\tilde{i}}\right)_{3,\a_{\tilde{i}}}^{3,\a_{\tilde{i}}}=b(\l,z_{\tilde{i}})$. Inserting the r.h.s. of (\ref{prp}) into Eq. (\ref{Tnn2}) one sees that the product $\prod_i\left(R_{0\tilde{i}}\right)_{3,\a_{\tilde{i}}}^{3,\a_{\tilde{i}}}$ creates a diagonal dressing factor for $\tilde{T}_{33}(\la)$ and the product of projectors applied to $R^{\s}(\la)$ gives $F_{1\ldots N}(\la)$. We thus obtain
\beqa
\tilde T_{33}(\l) = \otimes_{i=1}^N \mbox{diag}\{b(\l-z_i),b(\l-z_i),1\}\,.
\label{T33}
\eeqa\\
To compute $T_{32}(\l)$ one has to distinguish in the sum $\sum^*$ cases of various multiplicities $k_{3}$ and $k_{2}$ of the occurrence of group indices $3$ and $2$:
\beqa
&&\hspace{-2em}F_{1\ldots N}(\l)T_{3\,2}(\l)=
\sum_{\s\in {\cal{S}}_N}\sum_{k_{3}=0}^N\sum_{k_{2}=0}^{N-k_{3}}\sum^{\quad\quad *''}_{\a_{\tilde{1}}\ldots\a_{\tilde{N}}}\prod_{j_{3}=N-k_{3}+1}^N \hspace{-.5em}P_{\widetilde{j_{3}}}^{3}\prod_{j_{2}=N-k_{3}-k_{2}+1}^{N-k_{3}} \hspace{-.8em}P_{\widetilde{j_{2}}}^{2}\nonumber\\
&&\hspace{8em}\times\prod_{j=1}^{N-k_{3}-k_{2}} P_{\tilde{j}}^{\a_{\tilde{j}}} P^{3}_0 \;T_{0,\tilde{1}\ldots \tilde{N}} P^{2}_0
 R_{1\ldots N}^{\s}\,.\label{C11}
\eeqa
Evaluating the matrix product in $T_{0,\tilde{1}\ldots \tilde{N}}$ leads to
\beqa
&&\prod_{j_{3}=N-k_{3}+1}^N P_{\widetilde{j_{3}}}^{3}\prod_{j_{2}=N-k_{3}-k_{2}+1}^{N-k_{3}} P_{\widetilde{j_{2}}}^{2}\prod_{j=1}^{N-k_{3}-k_{2}} P_{\tilde{j}}^{\a_{\tilde{j}}} P^{3}_0\;T_{0,\tilde{1}\ldots \tilde{N}} P^{2}_0\nonumber\\
&=&\hspace{-1.2em}\sum_{i=N-k_{3}-k_{2}+1}^{N-k_{3}}\prod_{j=i+1}^{N-k_{3}}\left(R_{0\tilde{j}}\right)^{3\,2}_{3\,2}\left(R_{0\tilde{i}}\right)^{3\,2}_{2\,3}\prod_{N-k_{3}-k_{2}+1}^{j=i-1}\left(R_{0\tilde{j}}\right)^{2\,2}_{2\,2}\nonumber\\
&&\times\prod_{k=1}^{N-k_{3}-k_{2}}\left(R_{0\tilde{k}}\right)^{2\,\a_{\tilde{k}}}_{3\,\a_{\tilde{k}}}E^{(\tilde{i})}_{23}\prod_{j\neq i}P_{\tilde{j}}^{\a_{\tilde{j}}}P_{\tilde{i}}^{3}E^{(0)}_{32}\nonumber\\
&=&\hspace{-1.2em} \sum_{i=N-k_{3}-k_{2}+1}^{N-k_{3}}\prod_{l=i+1}^{N-k_{3}}b(\l,z_{\tilde{l}})c(\l,z_{\tilde{i}})\hspace{-.5em}\prod_{k=1}^{N-k_{3}-k_{2}}b(\l,z_{\tilde{k}})E^{(\tilde{i})}_{23}\prod_{j\neq i}P_{\tilde{j}}^{\a_{\tilde{j}}}P_{\tilde{i}}^{3}E^{(0)}_{32}\nonumber\\
\label{C12}
\eeqa
with $\left(E^{(i)}_{ab}\right)^{m}_{n}=\d_{m,a}\d_{n,b}$ denoting the root operators of $sl(3)$.\\
One notes that in the calculation the index $\a_{\tilde{i}}$ has changed from $2$ to $3$. As the distribution of $\a$'s is therefore no longer consistent with the conditions (\ref{cond}) in the sum $\sum^*$ one has to correct this fact by commuting the site $\tilde{i}$ through all higher sites $\tilde{j}$ with $\a_{\tilde{j}}=2$. So, taking into account (\ref{complaw}), one has to insert an additional factor $R_{\tilde{i}\,\widetilde{N-k_{3}}}\ldots R_{\tilde{i}\,\widetilde{i+1}}$
between the projectors and $R_{1\ldots N}^{\s}$ in (\ref{C11}). Because of Eq. (\ref{ppr}) no further corrections are necessary. For the following calculation we need
\beq
P^{3}_i\,P^{2}_j\,\id_{ij}=P^{3}_i\,P^{2}_j\left\{b(z_i,z_j)^{-1}R_{i\,j}-{{c(z_i,z_j)}\over{b(z_i,z_j)}}\Pi_{i\,j}\right\}\label{C1R}
\eeq
and
\beq
E^{(i)}_{3n}\,P^{3}_j\,\Pi_{i\,j}=E^j_{3n}\,P^{3}_i\,.\label{C1E}
\eeq
Let us now concentrate on the term with $i=N-k_{3}-k_{2}+1$ in (\ref{C12}) and use (\ref{C1R}) to create the needed factor above. Because of (\ref{C1E}) the second term in (\ref{C1R}) gives rise to an $E^{(\tilde{j})}_{23}$ with $\tilde{j}\neq \tilde{i}$. So the only possibility to get $E^{(\tilde{i})}_{23}$ is to use the first term in (\ref{C1R}). Corrections in the other terms with $j>N-k_{3}-k_{2}+1$ cannot lead to an expression with $E^{(\tilde{i})}_{23}$ as $\tilde{j}$ has not to be commuted with the site $\tilde{i}$. So the only term that contains $E^{(\tilde{i})}_{23}$ after the corrections for that special $R^{\s}_{1\ldots N}$ in (\ref{C12}) is
\beqa
\prod_{l=i+1}^{N-k_{3}}{{b(\l,z_{\tilde{l}})}\over{b(z_{\tilde{i}},z_{\tilde{l}})}}c(\l,z_{\tilde{i}})\prod_{k=1}^{N-k_{3}-k_{2}}b(\l-z_{\tilde{k}})E^{(\tilde{i})}_{23}.\label{C1sim}
\eeqa
Because of the symmetry of $\tilde{T}_{0,1\ldots N}(\l)$ all other terms have to be of the same form as (\ref{C1sim}). Taking into account the action of the projectors the resulting expression is
\beqa
{\tilde T}_{32}=\sum_{i=1}^N c(\l,z_i) E_{23}^{(i)}\otimes_{j\neq i}
\mbox{diag}\{
b(\l,z_j),b(\l,z_j)b^{-1}_{ij},1\}_{[j]}\;.\label{T_32}
\eeqa
For the calculation of $\tilde{T}_{3\,1}(\l)$ one has to distinguish the cases $3$, $2$ and $1$ in the sum $\sum^*$. The only difference compared to $\tilde{T}_{3\,2}(\l)$ is a term containing a product $E^{(i)}_{12}\,E^{(j)}_{23}$ now showing up in the matrix product in $T_{0,\s(1)\ldots \s{N}}(\la)$. Once again one has to correct the distribution of $\a's$ with the analog of the equations (\ref{C1R}) and (\ref{C1E})  
and also with a new relation which has to be taken into account when dealing with the term containing the product $E^{(i)}_{12}\,E^{(j)}_{23}$ :
$\;\;E^{(i)}_{13}\,P^{2}_j\,\Pi_{i\,j}=E^{(i)}_{12}E^{(j)}_{23}$.\\
This reasoning leads to ($b_{i\,k}=b(z_i,z_k),\;b_{0\,k}=b(\la,z_k)$):
\beqa
{\tilde T}_{31}\;&=&
\;\sum_{i=1}^N c_{0i} E_{13}^{(i)}\otimes_{j\neq i}
\mbox{diag}\{
b_{0j}b^{-1}_{ij},b_{0j}b^{-1}_{ij},1\}_{[j]}\;\,+\nonumber\\
\hspace{-1.5em}\sum_{i\neq j} c_{0i}&&\hspace{-1.5em}b_{0j}\,{{\eta}\over{z_i-z_j}}E_{23}^{(i)}\otimes
E_{12}^{(j)}
\otimes_{k\neq i,j}
\mbox{diag}\{
b_{0k}b^{-1}_{jk},b_{0k}b^{-1}_{ik},1\}_{[k]}.\label{T_31}
\eeqa
We have thus achieved to compute the creation operators in the new basis. They can be regarded as the corresponding Gaudin operators (defined as the semiclassical limit of the quantum monodromy matrix) being dressed diagonally.
\vspace{-.4cm}\section{Bethe Vectors}\setcounter{equation}{0}\vspace{-.1cm}
We now want to determine the functional form of the Bethe vectors (\ref{nbv}), using the explicit form of the relevant operators in the F-basis. First we want to note that the vacuum $\Omega^{(3)}_N$ (\ref{vac}) is invariant under the  F-transformation $F\,\Omega^{(3)}_N=\Omega^{(3)}_N$, as $R_{ij}v_3^{(i)}\otimes v_3^{(j)}=v_3^{(i)}\otimes v_3^{(j)}$, and thus due to the projectors in $F$ only the term containing the identity ($\{\a_1,\ldots,\a_N\}=\{3,\ldots,3\}$) contributes to $F\,\Omega^{(n)}_N$. The wavefunction (\ref{nbv}) in the F-basis is then $\tilde \Psi=F\Psi$:
\beq
\tilde \Psi_3 (N;\l_1,\ldots,\l_p)=\sum_{\a_1,\ldots,\a_{p}}
\Phi^{(2)}_{\a_1\ldots\a_{p}}
\tilde C_{\a_1}(\l_1)\ldots \tilde C_{\a_{p}}(\l_{p})\,\Omega^{(3)}_N.
\label{nbvf}
\eeq
We shall consider the ``off-shell'' Bethe vector \cite{bab}, i.e. without imposing the Bethe ansatz equations on $\{\l_i\}$.\\
We first consider the $sl(2)$ case. The F-transformed creation operators with respect to $\Omega^{(2)}_N$ are \cite{ms}
\beq
{\tilde C}(\l)=\sum_{i=1}^N c(\l,z_i)\s_+^{(i)}
\otimes_{j\ne i}
{\left(\matrix{
b(\l,z_j)b_{ij}^{-1}&0\cr
0\quad\quad&1\cr
}\right)}_{[j]}\, .
\label{T21}
\eeq
Inserting this expression into the $sl(2)$ wavefunction (\ref{phi2}), we obtain
\beqa
&&\tilde \Psi_2(N;\l_1,\ldots,\l_p)=\tilde{C}(\l_1)\ldots \tilde{C}(\l_p)\,\Omega^{(2)}_N\nonumber\\
&=&\sum_{i_1<\ldots< i_{p}} B_{p}^{(2)}(\l_1,\ldots,\l_{p}|z_{i_1},\ldots,z_{i_{p}})
\s_+^{(i_1)}\ldots\s_+^{(i_{p})}\,\Omega^{(2)}_N\;.
\label{Psi_2}
\eeqa
The coefficients $B_{p}^{(2)}(\l_1,\ldots,\l_{p}|z_{i_1},\ldots,z_{i_{p}})$ can be calculated by taking into account the action of the spin operators $\s^{(i)}_+$ on the dressing $\mbox{diag}\{b(\l,z_j)b_{ij}^{-1},1\}$ in (\ref{T21}):
\beqa
B^{(2)}_p(\l_1,\ldots,\l_p|z_{1},\ldots,z_{p})=
\sum_{\s\in S_p}\prod_{m=1}^p c(\l_m,z_{\s(m)})
\prod_{l=m+1}^{p}{{b(\l_{m},z_{\s(l)})}\over{b(z_{\s(m)},z_{\s(l)})}}.\nonumber\\
\label{B_2}
\eeqa
The strategy in the $sl(3)$ case will rely on the symmetry of the wavevector under the exchange of arbitrary spectral parameters which enables us to concentrate on a particularily simple term in the sum (\ref{nbvf}), and the repeated use of the commutation relation (\ref{crc}), which written in components is
\beqa
\tilde C_{\a}(\l_1)\tilde C_{\b}(\l_2)={{1}\over{b(\l_2,\l_1)}} \tilde C_{\b}(\l_2)\tilde C_{\a}(\l_1)
-{{c(\l_2,\l_1)}\over{b(\l_2,\l_1)}}\tilde C_{\b}(\l_1)\tilde C_{\a}(\l_2)\, .\nonumber\\
\label{crc2}
\eeqa
These ideas lead us to propose the following form for the Bethe vector
\beqa
&&{\tilde\Psi}_3(N,\l_1,\ldots,\l_{p_0};\l_{p_0+1},\ldots,\l_{p_0+p_1})= \nonumber\\
&&\sum_{\s \in S_{p_0}} B^{(2)}_{p_1}(\l_{p_0+1},\ldots,\l_{p_0 +p_1}|
\l_{\s(1)},\ldots,\l_{\s(p_1)})\prod_{k=1}^{p_1}\prod_{l=p_1+1}^{p_0}
b(\l_{\s(l)},\l_{\s(k)})^{-1}\nonumber\\
&&\times\;{\tilde C}_{2}(\l_{\s(p_1 +1)})\ldots {\tilde C}_{2}(\l_{\s(p_0)})
{\tilde C}_{1}(\l_{\s(1)})\ldots {\tilde C}_{1}(\l_{\s(p_1)})\,\Omega^{(3)}_N\, .
\label{Psi_3a}
\eeqa
Consider a special term in the sum (\ref{nbvf}) of the form (which is motivated by the fact that the associated coefficient $\Phi$ is especially simple to compute, see below)\,:
\beqa
{\tilde C}_{1}(\l_1)\ldots {\tilde C}_{1}(\l_{p_1})
{\tilde C}_{2}(\l_{p_1+1})\ldots {\tilde C}_{2}(\l_{p_0})\,\Omega_N^{(3)}\,.\label{12}
\eeqa
Commuting all ${\tilde C}_{1}(\l)$ to the right using the first term in (\ref{crc2}) (as the second terms produces expressions in which at least one  parameter of the operators $\tilde{T}_{31}$ does not belong to the specified set $\{\la_1,\ldots,\la_{p_1}\}$), yields an additional factor $\prod_{x=1}^{p_1}\prod_{y=p_1+1}^{p_0}\left\{b(\l_{y},\l_{x})\right\}^{-1}$.\\
It has to be noted that the associated c-number coefficient  $\Phi^{(2)}_{1\ldots 12\ldots 2}$ in (\ref{nbvf}) is not evaluated in the $sl(3)$ F-basis. It can however be expressed in the form (\ref{B_2}) as it is invariant under the action of the $sl(2)$ F-matrix. This is due to the fact that it constitutes a component of the $sl(2)$ vector whose labels (a non-decreasing series of $\a_i$ with respect to the original ordering of sites $i$) correspond via (\ref{cond}) to the identity permutation in the definition of the F-matrix (\ref{F}).\\
Invoking the exchange symmetry we arrive thus at the formula (\ref{Psi_3a}).\\
The explicit form of the creation operators (\ref{T_32}, \ref{T_31}) allows to further simplify the wavevector (\ref{nbvf}). The second term in (\ref{T_31}) annhilates the vacuum, which was the reason for the special choice of order in (\ref{Psi_3a}). Taking into account the recpective action of the roots on the dressings, we finally obtain our main result
\beqan
&&\hspace{-.3cm}{\tilde\Psi}_3(N,\l_1,\ldots,\l_{p_0};\l_{p_0+1},\ldots,\l_{p_0+p_1})=\nonumber\\
&&\hspace{-.4cm}\sum_{{i_1<\ldots<i_{p_0}}} \hspace{-.35cm}B_{p_0,p_1}^{(3)}(\l_1,\ldots,\l_{p_0};\l_{p_0+1},\ldots,
\l_{p_0+p_1}|z_{i_1},\ldots,z_{i_{p_0}})\hspace{-1em}\prod_{j=i_{p_1+1}}^{i_{p_0}}\hspace{-1em}E_{23}^{(j)}\prod_{j=i_{1}}^{i_{p_1}}\hspace{-.5em}E_{13}^{(j)}\,\Omega^{(3)}_N,
\eeqan
with
\beqa
&&\hspace{-.3cm}B_{p_0,p_1}^{(3)}(\l_1,\ldots,\l_{p_0};\l_{p_0+1},\ldots,
\l_{p_0+p_1}|z_{i_{1}},\ldots,z_{i_{p_0}})=\nonumber\\
&&\hspace{-.4cm}\sum_{\s \in S_{p_0}}\prod_{k=1}^{p_1}\prod_{l=p_1+1}^{p_0}{{b(\l_{\s(l)}-z_{i_k})}\over{b(\l_{\s(l)}-\l_{\s(k)})}}B_{p_0-p_1}^{(2)}(\l_{\s(p_1+1)},\ldots,\l_{\s(p_0)}|z_{i_{p_1+1}},\ldots,z_{i_{p_0}})\nonumber\\
&&\hspace{-.4cm}\times B_{p_1}^{(2)}(\l_{p_0+1},\ldots,\l_{p_0+p_1}|\l_{\s(1)},\ldots,\l_{\s(p_1)})B_{p_1}^{(2)}(\l_{\s(1)},\ldots,\l_{\s(p_1)}|z_{i_{1}},\ldots,z_{i_{p_1}}).\nonumber\\
\label{B1}
\eeqa
By expressing the $sl(3)$ wavevector (\ref{nbv}) with the help of $sl(2)$ building blocks (\ref{B_2}) we have achieved a resolution of the nested hierarchy.
\vspace{-.4cm}\section{Conclusion}\vspace{-.1cm}
We accomplished the construction of a factorizing F-matrix for the rational $sl(n)$ model enabling one to construct completely symmetric creation operators which moreover are devoid of non-local effects from polarization clouds. These operators were used to resolve the intricacies of the nested structure of the Bethe vectors.\\
Our method does not rely on an $sl(n)$ invariance of the monodromy matrix. The only ingredients needed in our computation are the form of the R-matrix (\ref{R}), i.e. its structure, unitarity and the fact that it constitutes a representation of the permutation group, and the property (\ref{ppr}). Thus our findings directly yield the corresponding expression for the generalized XXZ-model. With the help of a vertex--face map it is even possible to find the F-basis for the XYZ-model (\cite{abfpr}, see also R. Poghossians lecture in this volume) and its multicomponent generalization, the Belavin model \cite{ar}.\\
We hope that these results might prove useful for the construction of formfactors starting from the microscopic level. For this purpose the generalization of the above procedure for the thermodynamical limit would be desireable.\\
In view of the similarities between the results obtained by the functional Bethe ansatz \cite{skly, terras} and those in the F-basis for the $sl(2)$ XXX and XXZ-model, it seems plausible that some version of the functional Bethe ansatz should be feasible both for the ellipic case (XYZ-model) as well as for the nested form.
\vspace*{.5cm}\\
{\bf{Acknowledgement:}}
The work reported in these notes was done in collaboration with H. Boos and R. Flume.\\
We thank the organizers of the workshop for the opportunity to present this lecture and for their kind hospitality. \\

\end{document}